\documentclass[10pt, conference, compsocconf]{IEEEtran}
\usepackage[cmex10]{amsmath}
\usepackage[dvips]{graphicx}
\usepackage[caption=false,font=footnotesize]{subfig}
\usepackage{psfrag}
\usepackage{url} % \url{my_url_here}.
\usepackage{cite}

\begin{document}

\title{Probabilistic spreading of information in a spatial network}

\author{
\IEEEauthorblockN{
Krzysztof Malarz\IEEEauthorrefmark{1},
Vikas Chandra\IEEEauthorrefmark{2},
Eve Mitleton-Kelly\IEEEauthorrefmark{2},
Krzysztof Ku{\l}akowski\IEEEauthorrefmark{1}
}

\IEEEauthorblockA{
\IEEEauthorrefmark{1}Faculty of Physics and Applied Computer Science,\\
AGH University of Science and Technology,\\
al. Mickiewicza 30, PL-30059 Cracow, Poland\\
Email: malarz@agh.edu.pl, kulakowski@novell.ftj.agh.edu.pl}
\IEEEauthorrefmark{2}LSE Complexity Group\\
London School of Economics\\
London (UK) WC 2A 2AE\\
Email: V.Chandra@lse.ac.uk, E.Mitleton-Kelly@lse.ac.uk
}

\maketitle

\begin{abstract}
Spread of information in a crowd is analysed in terms of directed percolation in two-dimensional spatial network. We investigate 
the case when the information transmitted can be incomplete or damaged. The results indicate that for small or moderate probability 
of errors, it is only the critical connectivity that varies with this probability, but the shape of the transmission velocity curve
remains unchanged in a wide range of the probability. The shape of the boundary between those already informed and those yet uninformed
becomes complex when the connectivity of agents is small.
\end{abstract}


% \fi

%% \IEEEpeerreviewmaketitle

%% ===========================================================================
\section{Introduction}
%% ===========================================================================
We consider the process of information spreading in a spatial network. Our motivation is twofold. First, on the contrary to the scale-free networks \cite{wat,lkd,dog,hand,pasat,mejn}, spatial networks are less investigated in social systems; a recent exception is \cite{exc}. Still, in numerous applications the spatial distribution of agents does matter. This is so in particular when the effectiveness of communication depends on the geometrical distance between agents, as it is with visual or voice communication. Second---and this is the aim of this work---is to take into account possible errors or inaccuracy of transmission. This is an essential difference between standard modelling of this kind \cite{fortu,odor} and our model. We assume that at each time step a message is sent by each agent already informed to his/her each as-yet-uninformed neighbour, in each case with probability $p$.

The probability that a particular information bit is ‘damaged’ along the way can be considered as reflecting the complexity of the information so that even when heard, it is completely absorbed by an agent and not passed along to the neighbour. It may also reflect the level of ‘trust’  between an agent and his neighbour so that information is deemed to be passed along only when the source is trusted. It is quite likely that in an emergency and life threatening situation, the value of $p$ would be in the region of 0.5 and in such a case the group is unlikely to arrive at a consensus on the course of action to be taken and the spatial network of informed agents will remain unconnected. It is important in such a case that an external source that can be trusted whether it be an authority figure such as the driver of the underground train or an AmI device be used to give instructions about options available for rescue. This would artificially increase the value of $p$ and consequently increase the velocity of connections. The results indicate that this $p$ does not have to be perfectly 1 and as long as the level of trust does not go below a certain critical level---such as the driver himself needing help but still in a frame of mind to think rationally---the group should be able to consider a variety of options relatively quickly. This is in fact reflected in real events such as the July 7th, 2005 London Underground bombings where passengers waited for instructions to come through and followed the instructions of the driver even when the driver was injured and not in a position to offer all the necessary help \cite{bomb}.

%% ===========================================================================
\section{Calculations}
%% ===========================================================================
The algorithm is constructed as follows. First, the spatial network is determined: positions of nodes are randomly selected as points 
on a $2\times 2$ square. For a given tolerance parameter $\mu$, a circle of radius $\mu$ is set around each node $i$, and the nodes $j$ within
the circle are linked to the node $i$. The matrix element $c(i,j)$ of the connectivity matrix is set to be one, otherwise it is zero.

The simulation of the process starts from some number of nodes at one side of the square, which are treated as informed. The probability that 
a given message is passed along each link is denoted $p$, and the probability that this message is damaged or blocked is $q=1-p$. Then, at each 
simulation step, the number $k(i)$ of informed neighbours is found for each node $i$ which represents an agent who is not informed yet. During 
each time step, each as-yet-uninformed node is informed with the probability
\begin{equation}
P(i;p)=1-(1-p)^{k(i)}.
\end{equation}
In other words, to remain uninformed at given time step is equivalent to not-to-be informed by each of $k$ neighbours.

We measure the velocity, i.e. the average number of nodes being informed during one time step. The parameters of the simulation are: the number of nodes $N$, the tolerance parameter $\mu$ and the probability $p$. However, it is clear that for a given and fixed area an increase of $N$ four times and a simultaneous decrease of $\mu$ two times should remain the system locally unchanged. We are interested not that much in $\mu$, but rather in the mean degree $\langle k\rangle$. Below some critical value $\langle k\rangle_c$, the system remains not connected and the information does not spread at all. In the Erd\"os-R\'enyi networks, this critical value $\langle k\rangle_c=1$; however, the spatial network is much more correlated, with the clustering coefficient $C$ close to 0.46. It is straightforward to expect that $k$ varies with the tolerance parameter as $\mu^2$. The degree distribution $P(k)$ of the spatial network is Poissonian, i.e.
\begin{equation}
P(k)=\exp({-\langle k\rangle})\cdot\frac{\langle k\rangle ^k}{k!}.
\end{equation}

%% ===========================================================================
\section{Results}
%% ===========================================================================

%% ---------------------------------------------------------------------------
\begin{figure}
\psfrag{n}{$n$}
\psfrag{t}{$t$}
\psfrag{mu2}{$\mu^2=$}
\centering{\includegraphics[width=0.5\textwidth]{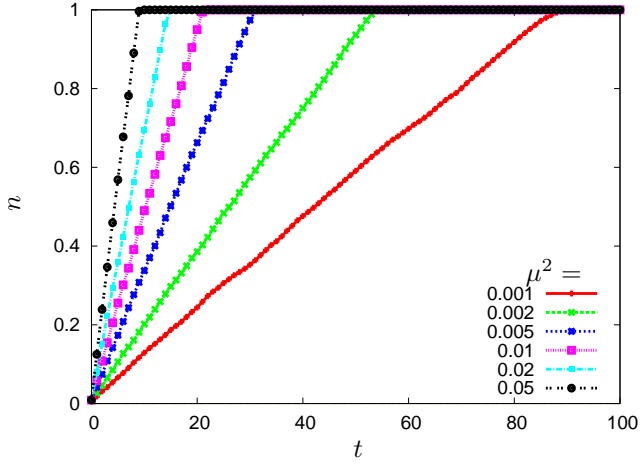}}
\caption{The number $n$ of informed nodes against time $t$ for different values of the tolerance parameter $\mu$. The velocity $\alpha$ in the next plot is the derivative of the plot presented here, before the saturation appears. $N=10^4$, $p=0.9$}
\label{f0}
\end{figure}
\begin{figure}
\psfrag{mu2}{$\mu^2$}
\psfrag{mu}{$\mu$}
\psfrag{avek}{$\langle k\rangle$}
\includegraphics[width=0.5\textwidth]{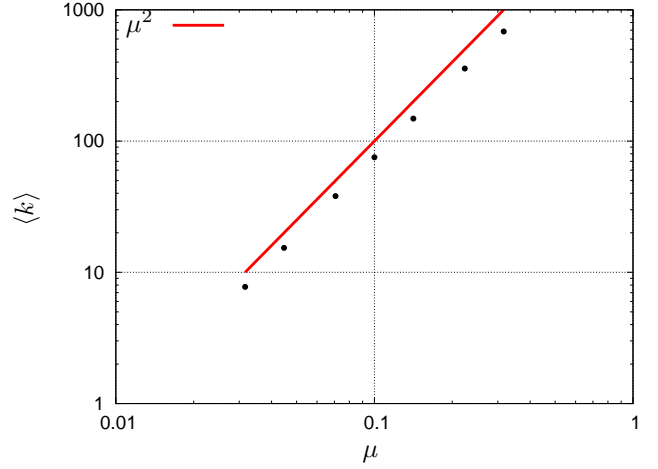}
\caption{The numerical proof that the mean degree $\langle k\rangle$ varies as $\mu^2$.}
\label{f2}
\end{figure}

%% ---------------------------------------------------------------------------
In Fig. \ref{f0} we show the number $n$ of informed agents against time $t$, measured in time steps defined above, for various values of the distance $\mu$ which allows for the communication. The relevant part of the plot is the ascending one; once the system boundaries are reached, $n$ does not 
increase any more. The inclination of the relevant part is the velocity $\alpha (\mu)$. The critical distance $\mu$ can be converted to the connectivity $\langle k\rangle$, i.e. the mean degree of the spatial network. As a consequence of the geometrical character of the 2-dimensional space, the mean connectivity $\langle k\rangle$ increases with $\mu$ as $\mu ^2$, as it is demonstrated numerically in Fig. \ref{f2}.

%% ---------------------------------------------------------------------------
%% ---------------------------------------------------------------------------

%% ---------------------------------------------------------------------------
\begin{figure}
\psfrag{alpha}{$\alpha$}
\psfrag{k-kkr}{$\langle k\rangle-k_c$}
\psfrag{p, kkr, beta}{$p$, $k_c$, $\beta$}
\includegraphics[width=0.5\textwidth]{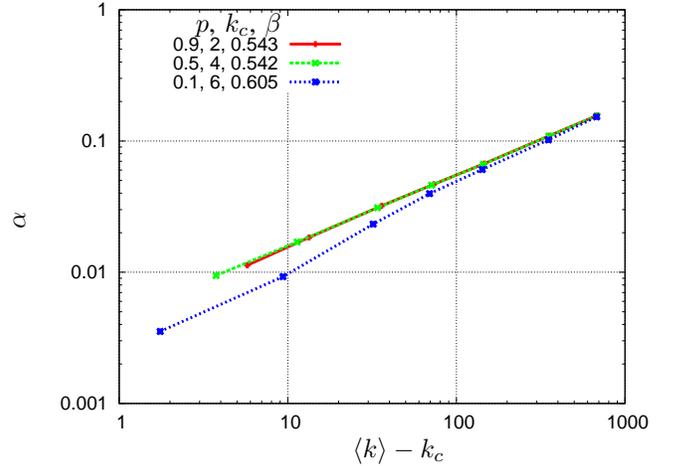}
\caption{The plot velocity $\alpha$ against $\langle k\rangle-k_c$ for three values of $p$.}
\label{f1}
\end{figure}
%% ---------------------------------------------------------------------------

%% ---------------------------------------------------------------------------
\begin{figure*}
\centering{
\subfloat[$N=10^4$, $p=0.1$, $\mu^2=0.02$]{\includegraphics[width=0.5\textwidth]{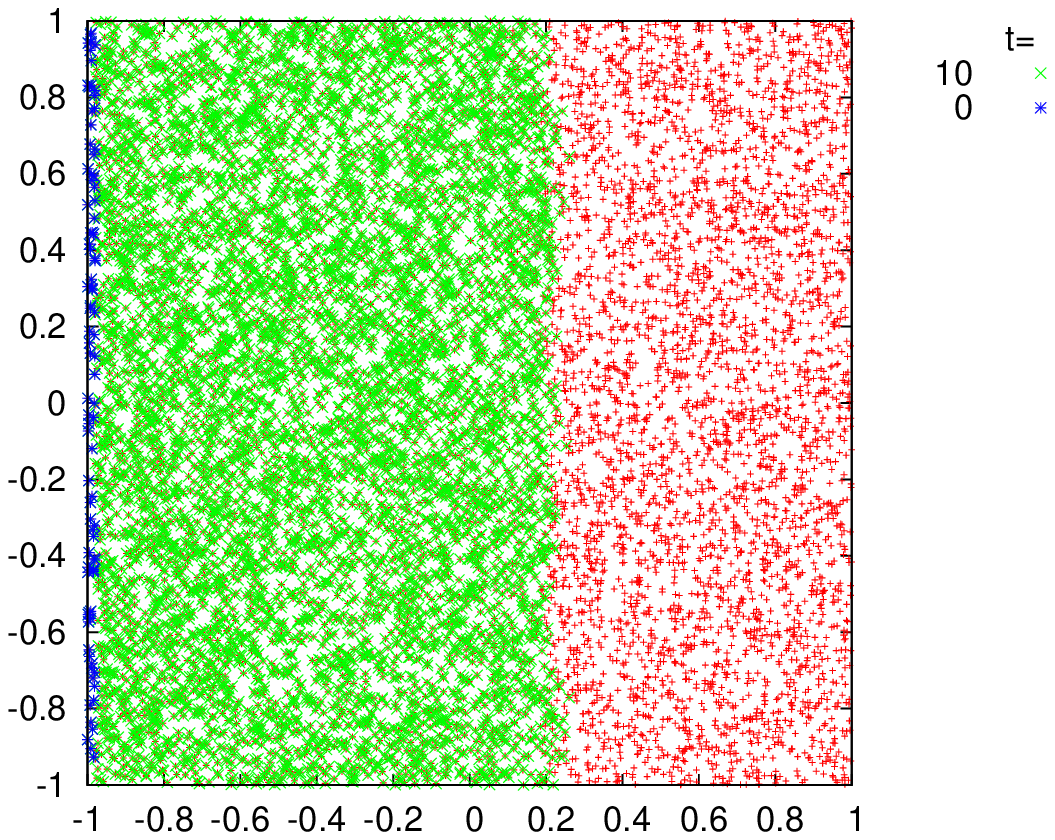}}\\
\subfloat[$N=10^4$, $p=0.1$, $\mu^2=0.001$]{\includegraphics[width=0.5\textwidth]{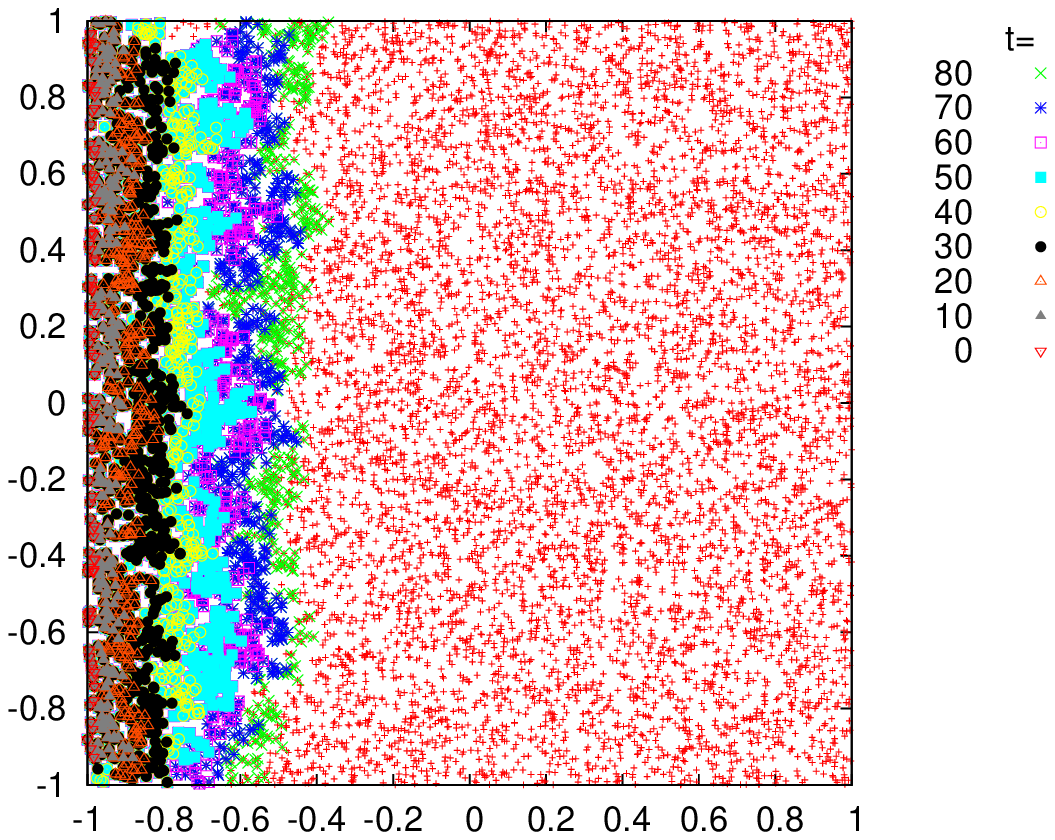}}
\subfloat[$N=10^4$, $p=0.1$, $\mu^2=0.0006$]{\includegraphics[width=0.5\textwidth]{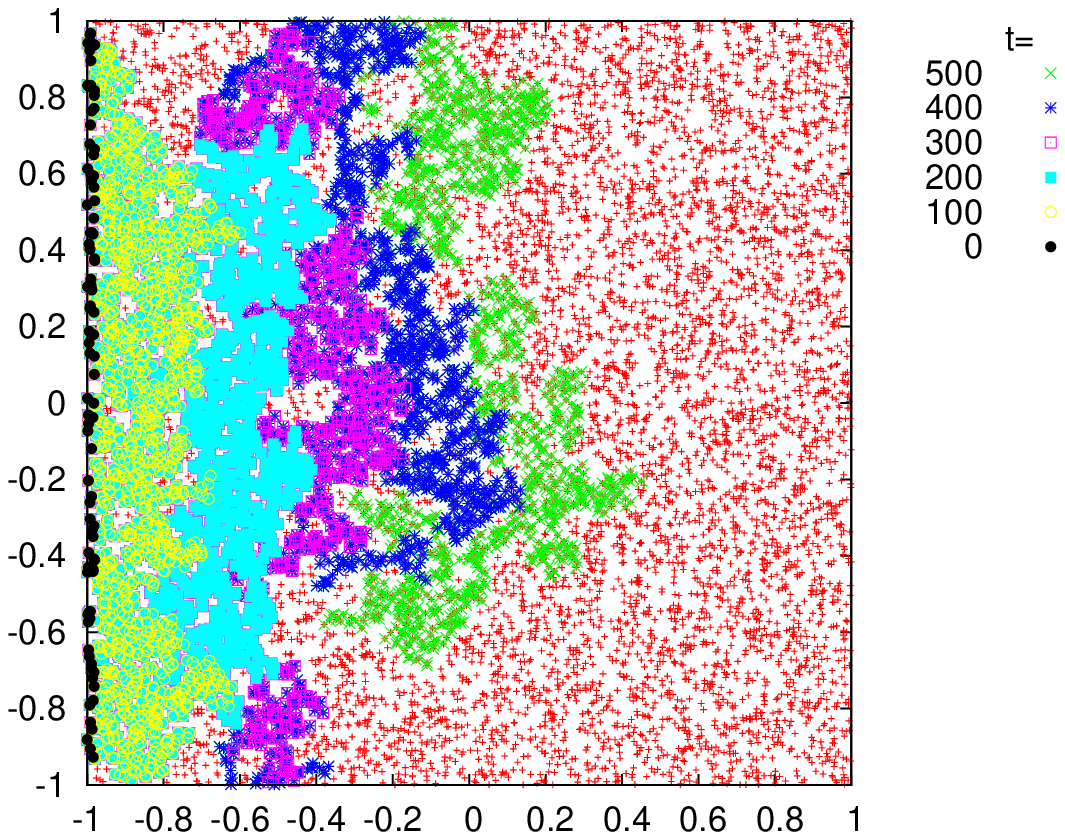}}\\
\subfloat[$N=10^4$, $p=0.9$, $\mu^2=0.001$]{\includegraphics[width=0.5\textwidth]{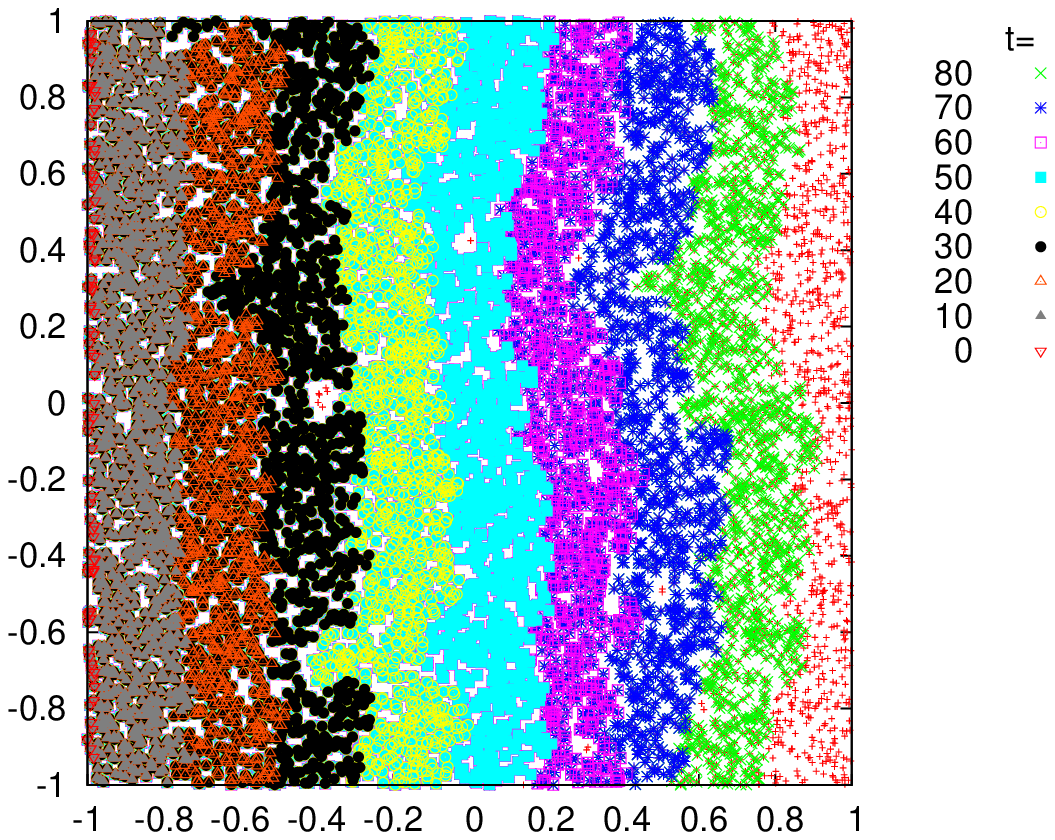}}
\subfloat[$N=10^4$, $p=0.9$, $\mu^2=0.0006$]{\includegraphics[width=0.5\textwidth]{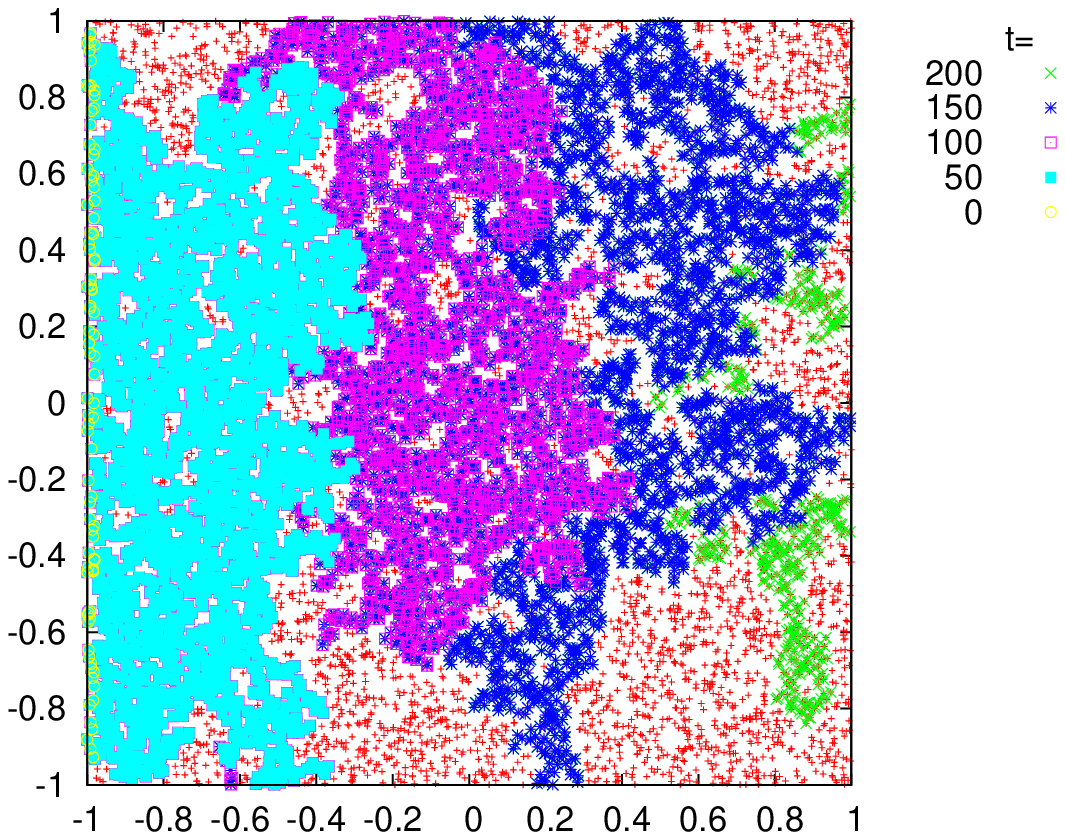}}
}
\caption{Visualisation of the front of being informed against time for various $\mu$.}
\label{f3}
\end{figure*}
%% ---------------------------------------------------------------------------

The velocity $\alpha$ is shown in Fig. \ref{f1}, as dependent on the difference between the connectivity $\langle k\rangle$ and the critical connectivity $k_c$.  
As we see, the plots for $p=0.9$ and $p=0.5$ coincide. We know from the data that also the plot for $p=1.0$ coincide with those two, with $k_c$=2 (the same as for $p=0.9$). However, the plot for $p=0.1$ does not coincide with the other ones. Still, the critical exponent $\beta$, defined as
\begin{equation}
\alpha \propto (\langle k\rangle-k_c)^\beta
\end{equation}
is almost the same for all investigated plots, and it is close to the value of two-dimensional directed percolation \cite{odor}, $\beta = 0.584$. We note that as in the vicinity of $k_c>0$ the connectivity $\langle k\rangle$ can be approximated by a linear function of $\mu$, the exponent $\beta$ is the same for 
the plots $\alpha (\langle k\rangle)$ and $\alpha (\mu)$.

In a series of subsequent plots \ref{f3} (a-e) we show the character of time and $p$ dependence of the boundary between the spatial areas occupied by informed and uninformed agents. As we see, the shape of the boundary is linear for large $\mu$, what means that the connectivity is also large. However, for small $\mu$ the boundary becomes a complex line, which reminds a fractal. We interpret this result as an interplay with the fluctuations of the connectivity $k$. In other words, the motion boundary can be stopped at some areas where agents are distributed with small density; which means that their number of neighbours is smaller, than the average value $\langle k\rangle$.

%% ===========================================================================
\section{Conclusion}
%% ===========================================================================

The aim of our calculations was to describe the possibility of incomplete or damaged message. The main result is that in these conditions, the only  modification of the results is the shift of the plot towards higher value of $k$. As we read from the legend, the critical connectivity $\langle k\rangle$ increases from 2 for $p=1$ and $0.9$ to about 6 for $p=0.1$. As remarked above, the plot for this value of $p$ deviates from the others. It seems to us however, that what is surprising here is not the deviation, but rather the coincidence of the plots $\alpha(\langle k\rangle-k_c)$ for $p=0.5$ and higher. This coincidence means that the low probability of the message transmission can be exactly compensated by the number of message senders.  In terms of statistical physics, the universality class of the transition seems to be the same for the deterministic ($p$=1) and the probabilistic ($p<$1) variant of the process; `seems to be', because the numerical evidence allows for only preliminary statements.

The results, found in a simple two-dimensional room, should apply also to other geometries, if only the connectivity distribution is not damaged by, for example, narrow corridors. What is specific for the spatial network is the lack of the small-world effect \cite{wat}. The spatial character of communication is specific for the visual or voice messages in a crowd, where both the spatial range and the probability of an efficient message transfer remain limited. A specific example of this problem is discussed elsewhere \cite{prz}. If the communication with electronic means is taken into account, this could not only accelerate the motion of the boundary shown in Fig. \ref{f3}, but also lead to an entire modification of the whole structure of the communication network. This problem will be discussed in a separate work.

A question appears, to what extent the results depend on our assumption is the only absorbing state in our model scheme: those uninformed can be informed and probably will be, while those informed probably will not forget. We could extend this irreversibility to both states, when we deal with spreading rather a decision to pass the message or not than a message itself. It is likely then, that an agent will find unfeasible to change his/her decision. In a limit case neither decision `No' nor `Yes' cannot be modified. When the amount of decisions `No' exceeds some critical value, the spreading is stopped. This variant seems to be analogous rather to the conventional percolation \cite{sta}, than to the directed one.

%% ===========================================================================
\section*{Acknowledgment}
%% ===========================================================================
The research is partially supported within the FP7 project SOCIONICAL, No. 231288.

%% \IEEEtriggeratref{12}

\end{document}